# Hooge's Constant for Carbon Nanotube Field Effect Transistors

Masa Ishigami, J. H. Chen, and E. D. Williams

Department of Physics, University of Maryland, College Park, College Park MD 20742

USA

David Tobias, Y. F. Chen, and M. S. Fuhrer

Department of Physics and Center for Superconductivity Research, University of

Maryland, College Park, College Park MD 20742 USA

**Abstract**

The 1/f noise in individual semiconducting carbon nanotubes (s-CNT) in a field-effect transistor configuration has been measured in ultra-high vacuum and following exposure to air. The amplitude of the normalized current spectral noise density is independent of source-drain current, and inversely proportional to gate voltage, to channel length and therefore to carrier number, indicating the noise is due to mobility rather than number fluctuations. Hooge's constant for s-CNT is found to be $9.3 \pm 0.4 \times 10^{-3}$. The magnitude of the 1/f noise is substantially decreased by exposing the devices to air.

Carbon nanotubes (CNTs) are now widely considered to be one of the most promising nanoscale materials for electronics applications[1]. Field effect transistors (FETs) fabricated using semiconducting CNTs (s-CNTs) have high carrier mobility[2] and high current capacity[3]. Memory devices[4,5], and chemical[6-8] and biological[9] sensors can be fabricated using s-CNT-FETs. However, the ratio of electronic noise to device signal is expected to increase in devices of decreasing size[10] and is thus of concern in nanoscale devices. In addition, surface adsorbates[11] and atomic scale structural fluctuations[12] are expected to have increased influence on electronic noise as the surface to volume ratio increases. There are previous reports on electronic noise in a metallic CNT[13], CNT ropes[14,15] and networks[14,16]. However, a quantitative rule for determining the noise amplitude is not yet known, mostly because the number of carriers could not be accurately determined in the previous work. Electronic noise in devices fabricated from individual s-CNTs, which are more technologically important, remains unexplored.

We have characterized[17] electronic noise in FETs fabricated from individual s-CNTs in ultra high vacuum in the "on" state of devices. We find that the gate-voltage and channel-length dependence of the amplitude of the $1/f$ noise is consistent with Hooge's empirical rule for noise caused by mobility fluctuations and not by number fluctuations. Furthermore, we also find that the $1/f$ noise decreases when the same devices are subsequently measured in air.

The s-CNTs used in our experiment were grown from iron-based catalysts[18] by chemical vapor deposition[19] on thermally-grown $SiO_2$ on degenerately-doped Si substrates. Electron beam lithography was used to define Cr/Au electrodes in a two-probe configuration with channel lengths of 1.6 to 28 μm. CNT devices were annealed in a $H_2$/Ar flow at 400 °C to improve contact resistance prior to transport measurements. All measurements were performed unless otherwise noted in an ultra high vacuum (UHV) environment, in order to measure intrinsic device noise devoid of any interference from adsorbates. Devices were degassed at 200 degrees for longer than one hour in UHV prior to noise measurements performed *in situ*. The transport characteristics are similar to those of individual s-CNT transistors measured in air[20].

We first examine the current (*I*) and frequency (*f*) dependence of the low-frequency noise. In the linear IV regime (described below), 1/*f* noise is expected to follow the relationship:

$$S_I = \frac{AI^2}{f}$$

In our experiment, we typically find $S_I \propto I^{2\pm0.1}$. The frequency dependence of the noise reveals two types of low frequency noise spectra, which are observed with an equal prevalence. An example of the first type of spectrum is shown in Fig. 1a. The inverse of the normalized noise power is proportional to frequency; i.e. electronic noise in the first type of spectrum is strictly 1/*f*. Fig. 1b shows an example of the second type of spectrum. Such a minor deviation from 1/*f* dependence can be due to random telegraph noise (RTS) often present in nanotube FETs[21] and sub-micron CMOS FETs. It is generated by trapping-detrapping of carriers by tunneling into traps in the $SiO_2$[4,21]. The deviation observed can be well explained by the addition of a generation-recombination (GR) noise term[22] which adequately describes RTS:

$$S_I = \frac{AI^2}{f} + \frac{BI^2}{1+(\frac{f}{f_0})^2},$$

where $f_0$ is the characteristic frequency for the GR noise. As shown in Fig. 1b, an accurate curve fit to the second type of spectra is made by adding a minor GR component to the $1/f$ component.

We now turn to the gate voltage dependence of the noise amplitude coefficient $A$ in the "on" state in the linear regime, defined by $V_g < V_{th}$ and $|V_d| << |V_g - V_{th}|$, where $V_d$, $V_g$, and $V_{th}$ are the drain voltage, gate voltage, and device threshold voltage respectively. For the case of mobility fluctuations, Hooge's empirical rule[10] states that the noise coefficient $A$ is given by

$$A = \frac{\alpha_H}{N},$$

where $\alpha_H$ is Hooge's constant and $N$ is the total number of carriers in the system. Since $N = c_g L |V_g - V_{th}|/e$ in a one-dimensional FET in the "on" state, the above equation may be rewritten as

$$\frac{1}{A} = \frac{c_g L |V_g - V_{th}|}{\alpha_H e}$$

where $L$ is the device length and $e$ is the electronic charge. The gate capacitance per length $c_g$ is given by[23]

$$c_g = \frac{2\pi \varepsilon_0 \varepsilon_{av}}{\ln\left(\frac{2z}{d}\right)}$$

where $\varepsilon_{av}$ is the average dielectric constant of the gate dielectric and the medium above the CNT, $z$ is the thickness of the gate dielectric, and $d$ is the diameter of CNTs.

There are two classes of models[24] for 1/f noise in metal-oxide-semiconductor field effect transistors (MOSFETs). Models based on mobility fluctuations predict that $\alpha_H$ is a independent of gate voltage, while $\alpha_H \propto 1/|V_g-V_{th}|$ in models based on number fluctuations. As such, in the linear regime, $1/A \propto |V_g-V_{th}|$ if noise is due to mobility fluctuations and $1/A \propto |V_g-V_{th}|^2$ if noise is due to number fluctuations.

Fig. 2a shows the transfer characteristics of a typical (p-type) s-CNT FET. This representative FET has a turn-on voltage or threshold gate voltage of -8.6 volts. We find that $1/A$ is linearly dependent ($V^{1.08\pm0.09}$) on the gate voltage and has a zero intercept at the threshold gate voltage as shown in Fig. 2b. Thus, the gate dependence of $A$ is consistent with Hooge's rule with mobility fluctuations, i.e. $1/A = D|V_g - V_{th}|$, where $D$ is a constant, equal to $c_g L/\alpha_H e$. We have found the constant $D$ for 6 device lengths, all with 500 nm thick gate oxide; the measured diameters range from ~1.2-1.8 nm for the CNTs in the devices. Fig. 3 shows $De/c_g$ as a function of length $L$. Consistent with the dependence on the number of carriers in Hooge's empirical rule, $De/c_g$ is proportional to $L$, with a slope $1/\alpha_H = 106.5\pm4.0$ ($\alpha_H = 9.3\pm0.4 \times 10^{-3}$). We note that the linear dependence of $D$ on length indicates that the noise arises due to fluctuations of a resistance that is proportional to the CNT length. This resistance must be the diffusive resistance of the CNT itself and not the contact resistance, which is independent of $L$.

The measured Hooge's constant is much smaller than $\alpha_H = 0.2$ previously reported[14] for CNT ropes and mats, for which the number of carriers is ambiguous due to mixture of metallic and semiconducting nanotubes. The previous report[14] attributed the excess noise to surface adsorbates. Quite interestingly, we find that the noise amplitude coefficient $A$ is significantly smaller in air than in UHV as shown in Fig. 4. The slope of $1/A$ vs. $V_g$ is typically larger by a factor of ~2-4 in air compared to UHV. However, it is unclear whether this apparent quieting of the noise is due to a reduction of the noise generation process in air, or an increase in $c_g$ due to the surface-adsorbed water layer (with its high dielectric constant). Correlating noise with controlled adsorption under vacuum will be needed to address this question.

In conclusion, we find that Hooge's empirical rule adequately describes the low-frequency noise in s-CNT FETs with $\alpha_H = 9.3\pm0.4\times10^{-3}$. The amplitude of 1/$f$ noise is inversely proportional to $|V_g-V_{th}|$ indicating mobility fluctuations and ruling out number fluctuations as the cause. The amplitude is also inversely proportional to the device length demonstrating that the noise is a property of the length-dependent resistance of the CNT and not the electronic contacts. The value for Hooge's constant is derived from the length dependence and is within the range of values observed in conventional three-dimensional semiconductors. This agreement in notable, given that, to our knowledge, $\alpha_H$ has not been measured in a one-dimensional semiconductor. Furthermore, we have shown that surface adsorbates due to air exposure do not increase the magnitude of the 1/$f$ noise, contrary to expectations[14]. This letter lays to rest any concerns about anomalously large noise in CNTs.

Note: While writing our manuscript, we became aware of a concurrent work reported in a preprint posted by Lin et al. (arXiv:cond-mat/0512595).

Acknowledgements: This work has been supported by the Director of Central Intelligence Postdoctoral Fellowship, the Laboratory for Physical Sciences, and NSF Grant No. 01-02950. The UMD NSF-MRSEC shared experimental facilities were used in this work.

**Figure Captions**

Figure 1

(a) Inverse of normalized noise power ($I^2/S_I$) of a s-CNT FET. The spectrum was acquired with source-drain bias at 10 mV, device current at 136 nA, and gate voltage at -7.5 V. The threshold voltage for the transistor was -2.5 V. The inverse noise power varies linearly with respect to frequency as shown by the solid line indicating the low frequency noise is $1/f$. (b) Inverse of normalized noise power ($I^2/S_I$) at different bias voltage of a second s-CNT FET. The solid line is a fit to the spectrum with both $1/f$ and generation-recombination noise terms (described in text) with an A/B ratio of 790.

Figure 2

(a) Conductance vs. gate voltage ($G$-$V_g$) of a s-CNT device with length of 5 μm. Conductance is expressed in terms of the conductance quantum $G_o = 2e^2/h = 77.5$ μS. (b) $1/A$ or $S_I f / I^2$ plotted against $V_g$. The solid red line is a linear fit to the data excluding the two points at $V_g > -8.8$ volt. Such linear dependence is typically seen for $|V_g - V_{th}| < 6\text{-}10$ volts.

Figure 3

Slope extracted from $1/A$ vs. $V_g$ data (Figure 3) in units of $c_g/e$ for six s-CNT devices with 500 nm thick gate oxide plotted with respect to the device lengths. The solid red line is a linear fit for the data; the slope is 106.5±4.0 for the line.

Figure 4

Comparison of inverse noise amplitude $1/A$ versus gate voltage $V_g - V_{th}$ for the same s-CNT device in UHV and in air. The amplitude of the $1/f$ noise in air is 3 times smaller than in UHV.

Figure 1

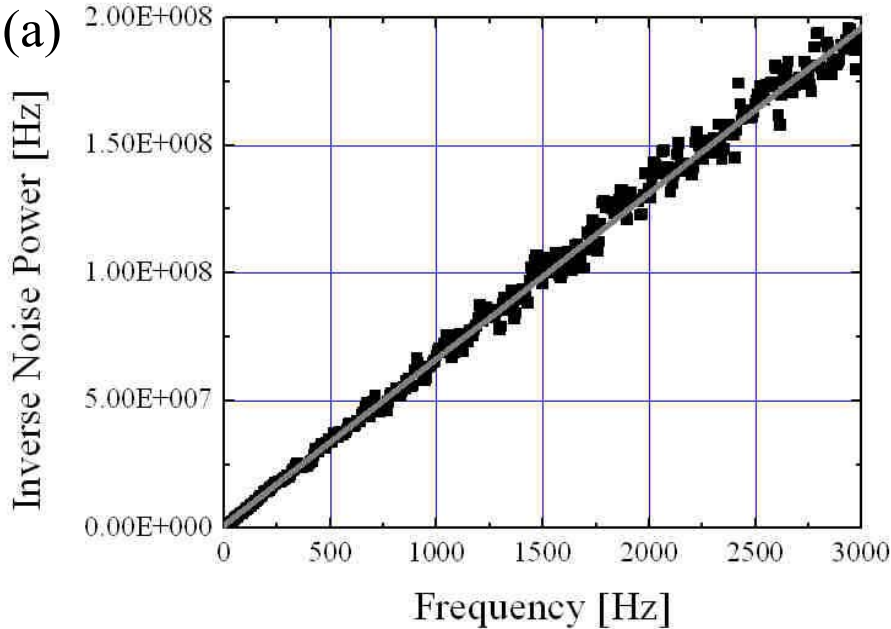 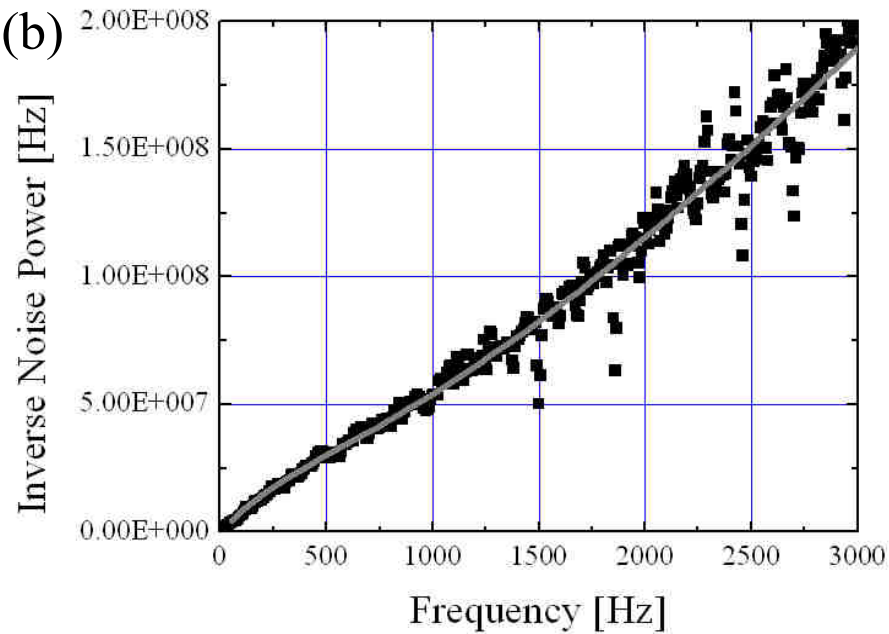

Figure 2

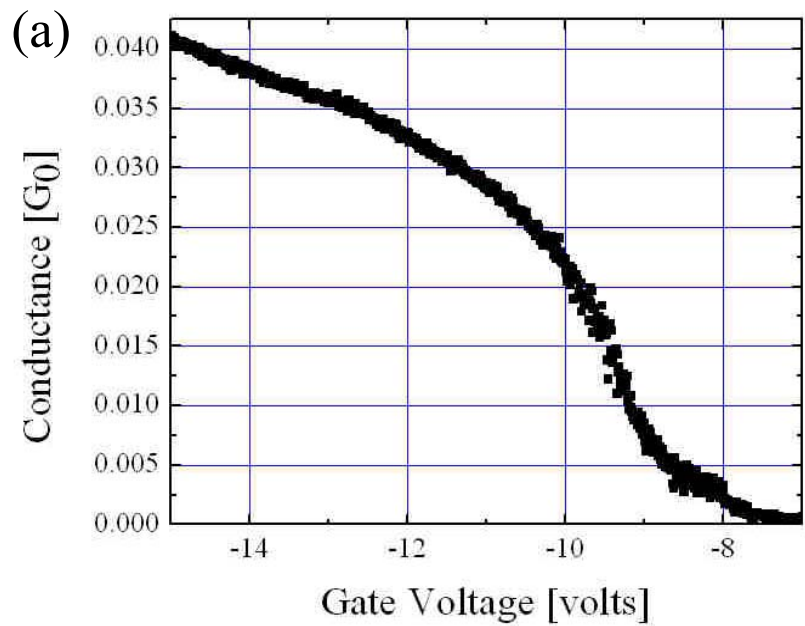 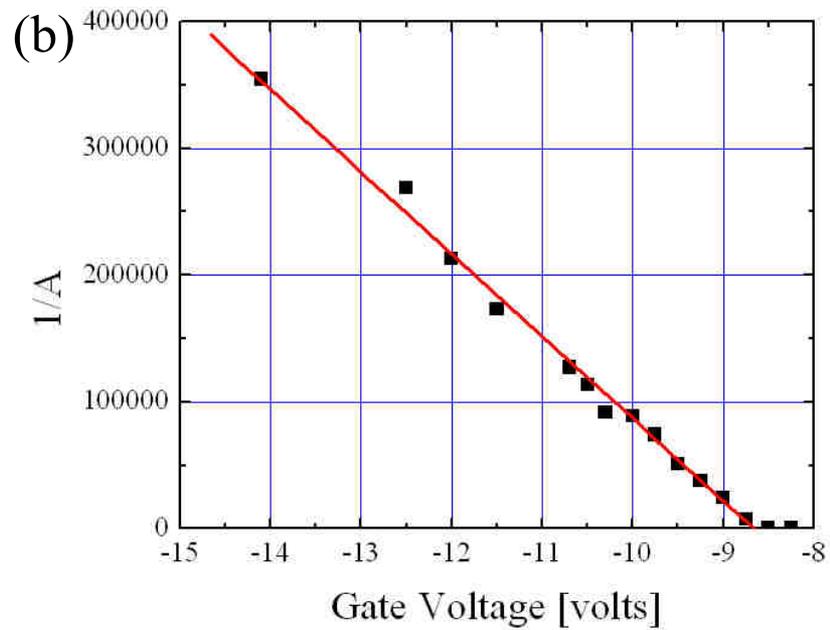

Figure 3

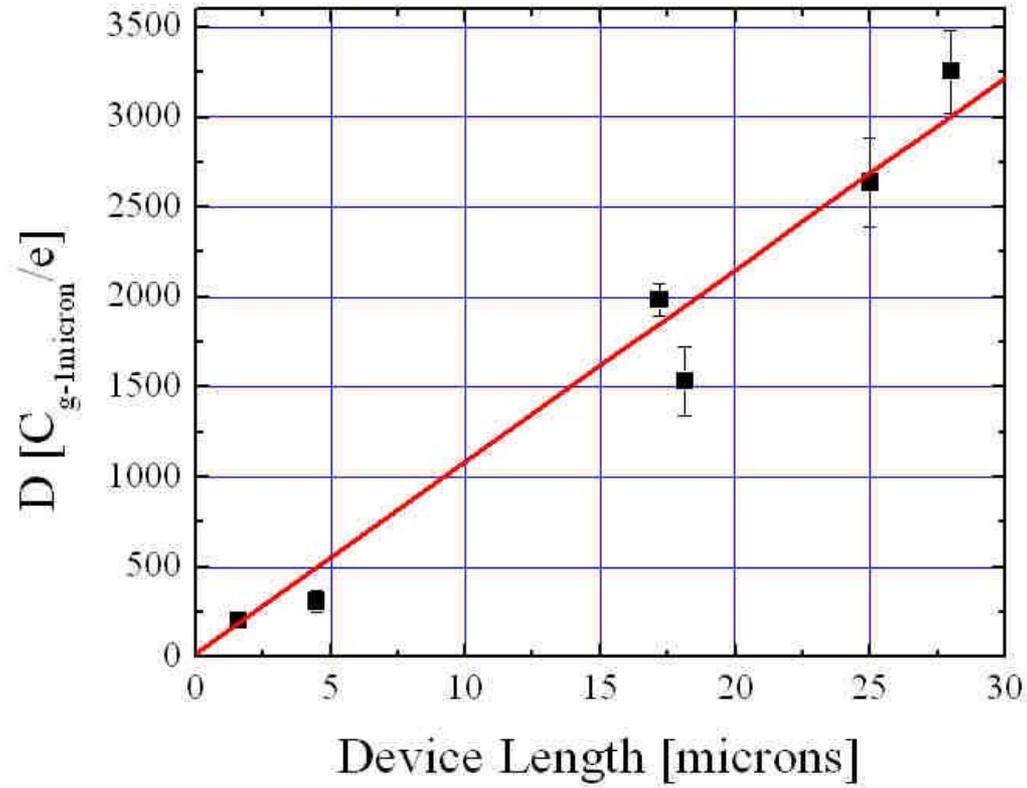

Figure 4

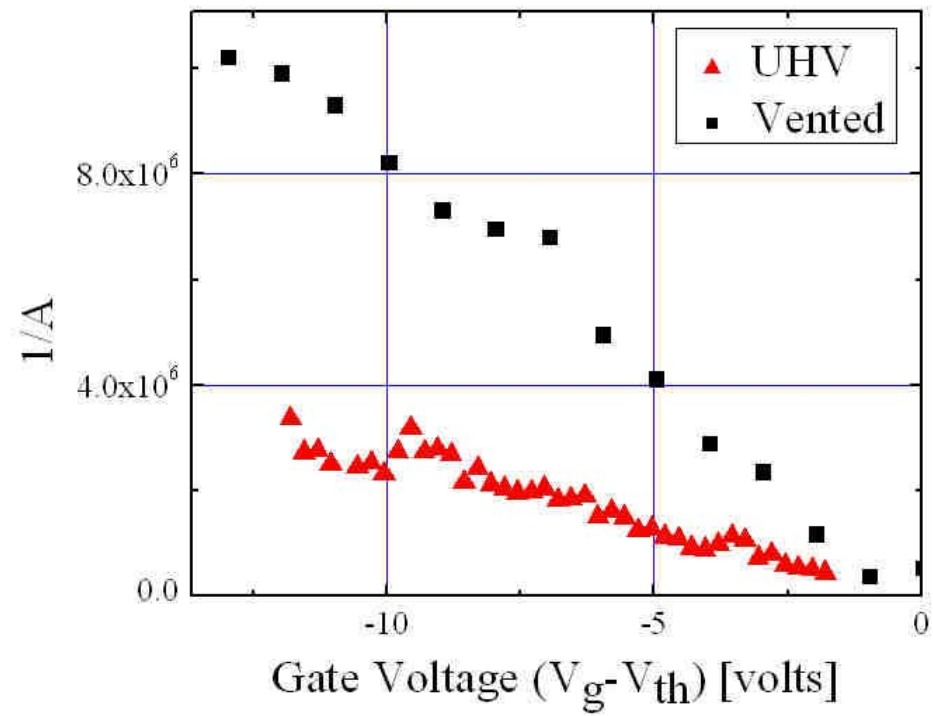